\documentstyle[psfrag,aps,preprint,epsfig,axodraw]{revtex}

\begin{document}

\tighten

\preprint{TU-700}
\title{New Constraint on Squark Flavor Mixing \\
from $^{199}$Hg Electric Dipole Moment}
\author{Motoi Endo\footnote{e-mail: endo@tuhep.phys.tohoku.ac.jp}, 
Mitsuru Kakizaki\footnote{e-mail: kakizaki@tuhep.phys.tohoku.ac.jp} and 
Masahiro Yamaguchi\footnote{e-mail: yama@tuhep.phys.tohoku.ac.jp}}
\address{Department of Physics, Tohoku University,
Sendai 980-8578, Japan}
\date{\today}
\maketitle
\begin{abstract} 
  A new constraint is obtained on the CP-violating flavor mixing
  between the left-handed top scalar quark $(\tilde{t}_L)$ and charm
  scalar quark $(\tilde{c}_L)$, by considering a chargino loop
  contribution to chromo-electric dipole moment of strange quark,
  which is limited by the electric dipole moment of the $^{199}$Hg
  atom. It is found that the flavor mixing should be suppressed to the
  level of ${\cal O}(0.1)$ for the  CP phase of order unity, 
  when $\tan \beta$ is relatively large and
  sparticle masses lie in a few hundred GeV range. Although it is much
  stronger than the known constraint from the chargino loop
  contribution to $b \to s \gamma$, the moderate constraint we obtain
  here is argued to leave room for sizable supersymmetric
  contribution to the  CP asymmetry in $B_d^0 \to \phi K_s$.
\end{abstract} 

\clearpage

Although low energy supersymmetry (SUSY) is thought to be one of the
most promising solutions to the naturalness problem on the electroweak
scale inherent in the standard model \cite{susy}, a random choice of
supersymmetry breaking parameters would lead to phenomenological
disaster.  Contributions from superparticle loops would induce too
large flavor changing neutral currents (FCNCs) \cite{flavor-susy}. 
Supersymmetric contributions to electric dipole moments
(EDMs) would also exceed experimental upper bounds \cite{CP-susy}. 
These two problems are referred to as the SUSY flavor
problem and the SUSY CP problem, respectively.

Supersymmetry breaking and its mediation to the standard model sector
must therefore be well controlled to avoid the aforementioned disaster.
One way to escape from the SUSY flavor problem is to invoke
hypothetical degeneracy of masses of squarks and sleptons among
different generations.  This approach includes minimal supergravity 
\cite{mSUGRA},
gauge mediation \cite{gauge-mediation}, anomaly mediation 
\cite{anomaly-mediation} and gaugino mediation \cite{IKYY,gaugino2}. Another
way is an alignment mechanism \cite{alignment,Dine:1993np,Pouliot:1993zm,Kaplan:1993ej,Pomarol:1995xc,Barbieri:1995uv,Hall:1995es,Hamaguchi:2002vi} 
 of generation mixing of sfermion mass terms
with their fermion counterparts. 
This requires understanding not only sfermion mass generation
but also fermion mass generation mechanism. Though appealing, it is rather
non-trivial to construct convincing models of flavor. 

To see what kind of mediation mechanism really works, it is important
to explore possible constraints coming from experimental data. As for
the constraints on the mixing of the first two generations, this issue
has been studied extensively in the literature (see Ref. 
\cite{Gabbiani:1996hi} and references therein).  
On the other hand, the constraints involving the third generation are still
under study and this is the subject of the present paper.

A remarkable observation has recently been made by Hisano and Shimizu,
who have pointed out that supersymmetric contribution to the
chromo-electric dipole moment (CEDM) of the strange quark is strongly
constrained by the EDM of the $^{199}$Hg atom \cite{Hisano:2003iw}. In
particular they have found that a product of left- and right-handed
squark mixings suffers from a stringent constraint.
  This is very interesting because the flavor
conserving process (EDM) constrains the flavor mixing of
squarks. Moreover the tight constraint on the squark flavor mixings
puts an important impact on SUSY contribution to the CP asymmetry in
$B_d^0 \to \phi K_s$.  In fact, the bound implies that the
contribution from $\tilde{b}_R - \tilde{s}_R$ mixing should be
extremely small, giving a severe restriction on an interesting
scenario where large contribution comes from renormalization group
effects of right-handed neutrino Yukawa interaction in grand unified
SUSY models \cite{Moroi:2000tk}.

In the present paper, we shall point out that the bound on the EDM of
the $^{199}$Hg atom  gives a new constraint on the up-type
left-handed squark mixing  when one considers chargino-loop
processes.  We will compare this constraint with that from $b \to
s \gamma$ and also argue implications to the CP asymmetry in  
$B_d^0 \to \phi K_s$.

Let us first clarify some notations on the squark masses and mixing used in the
literature \cite{Hall:1985dx,Gabbiani:1996hi}. 
It is convenient to define
\begin{eqnarray}
  & & (\delta^u_{LL})_{ij}=\frac{(m^2_{\tilde u_L})_{ij}}{m^2_{\tilde q}}, 
  \quad
  (\delta^u_{RR})_{ij}=\frac{(m^2_{\tilde u_R})_{ij}}{m^2_{\tilde q}},
  \nonumber \\
  & & (\delta^u_{LR})_{ij} 
  = - \frac{(m^2_{\tilde uLR})_{ij}}{m^2_{\tilde{q}}}, \quad
  \quad (\delta^u_{RL})_{ij} = (\delta^{u*}_{LR})_{ji}
\end{eqnarray}
where $(m^2_{\tilde u_{L(R)}})_{ij}$ is an $(ij)$ element of the
left-handed (right-handed) squark mass squared matrix for up-type squarks
in the superCKM basis, 
in which the Yukawa coupling matrices are in the diagonal form,
and $m_{\tilde q}$ is an average squark mass. 
A similar notation will be used for the down-type squarks.

Next we would like to briefly review the CEDM for quarks and its 
contribution to the EDM of the $^{199}$Hg atom. We
follow  Ref.~\cite{Falk:1999tm}.
The effective Hamiltonian for the CEDM for quarks is written
\begin{equation}
    H=\sum_{q=u,d,s} d_q^C \frac{i}{2} 
    g_s \bar{q} \sigma^{\mu \nu} T^A \gamma^5 q G_{\mu \nu}^A,
    \label{eq:CEDM_operator}
\end{equation}
where $g_s$ is the strong coupling constant, $G_{\mu \nu}^A$
represents the gluon field strength with $SU(3)_C$ generator index $A$.
The CEDM of the quark $q$ is denoted by $d_q^C$. 
The EDM of the $^{199}$Hg
atom is dominated by the T-odd nuclear force in $\pi^0$ and $\eta$
couplings to nucleons, which is generated by the CEDMs of the
constituent quarks. The EDM of the mercury atom was evaluated as
\begin{equation}
  d_{Hg}=-e (d_d^C-d_u^C-0.012 d_s^C) \times 3.2 \times 10^{-2},
\end{equation}
which is constrained by the existing experimental data,
$|d_{Hg}|<2.1 \times 10^{-28}\ e\ \mbox{cm}$ at 95\% C.L. 
\cite{Romalis:2000mg}.
Despite the small numerical coefficient in front of $d_s^C$, the strange
quark contribution can be important since $d_s^C$ itself is enhanced by
a heavier quark mass in the second (and sometimes third) generation.
When the contributions from the up and down quarks are absent, the constraint
on the EDM of the atom 
gives the bound to the CEDM of the strange quark as
\begin{equation}
    e|d_s^C| < 5.5 \times 10^{-25}\ e\ \mbox{cm}.
\label{eq:bound-on-cedm-of-s} 
\end{equation}
The chromo-dipole moment 
may also be constrained by the neutron EDM.
However the computation of the latter is more model-dependent\footnote{
A parton quark model is found to give a severer constraint 
on the squark flavor mixing from strange quark contribution compared to 
that from the $^{199}$Hg atom.  It is argued,however, that it is likely to be 
an overestimate \cite{Falk:1999tm,Abel:2001vy}. 
More conservative estimation shows the neutron EDM provides a similar limit 
with the $^{199}$Hg EDM, though it is model-dependent.}
and thus we do not consider the constraint coming from the neutron EDM.

Since EDMs are flavor conserving processes,
contributions from the (flavor conserving) CP phases of $A$ and $B$ 
parameters have been 
extensively discussed in the framework of SUSY models.
By imposing the experimental  bounds of 
the EDMs of the electron, neutron and mercury atom
on EDM operators generated by flavor diagonal interactions,
we obtain
$\mbox{Im}\ A \lesssim 10^{-2} - 10^{-1}$ and 
$\mbox{Im}\ B \lesssim 10^{-3} - 10^{-2}$ \cite{Abel:2001vy}.
Here we do not discuss possible mechanisms to suppress the CP phases of 
the $A$ and $B$ parameters, and simply assume that they satisfy 
the bounds given above.

On the other hand,  Hisano and Shimizu have found  that the
bound Eq.~(\ref{eq:bound-on-cedm-of-s}) gives a very stringent
constraint on squark flavor mixing \cite{Hisano:2003iw}. 
Specifically they have pointed out
that the CP violating part of the product of $(\delta^d_{LL})_{23}$
and $(\delta^d_{RR})_{32}$ is strongly constrained.  
The reason is that in the presence of sizable
$\tilde{b}_L-\tilde{s}_L$ and $\tilde{b}_R-\tilde{s}_R$ mixing 
the dominant contribution to the CEDM of the strange quark arises from 
the gluino-loop diagram with the $\tilde{b}_L-\tilde{b}_R$ mixing, 
which is proportional to $m_b \mu \tan \beta$,
and thus is enhanced by $m_b/m_s$ over the usual contribution induced by 
the $\tilde{s}_L-\tilde{s}_R$ mixing.

Although their constraint is only for the product of
$(\delta^d_{LL})_{23}$ and $(\delta^d_{RR})_{32}$, it practically
restricts the phase of $(\delta^d_{RR})_{32}$ in most models with high
energy SUSY breaking. This is because $(\delta^d_{LL})_{23}$ is
expected to emerge inevitably at low energy scales due to
renormalization group effects even if it is absent at the cutoff
scale, say, the Planck scale.  At the electroweak scale it is
evaluated as $(\delta^d_{LL})_{23} \sim 0.01$ at least unless there is
accidental cancellation.  Therefore the bound of the $^{199}$Hg EDM
constrains the $2-3$ mixing in the right-right (RR) sector
as $|\mbox{Im} (\delta^d_{RR})_{32}| < O(10^{-3})$.  This has an
important implication to $B_d^0 \to \phi K_s$. The Wilson coefficient
from the gluino penguin diagrams which contribute to the $B_d^0 \to
\phi K_s$ strongly correlates with that for the $^{199}$Hg EDM.
  As a result, the CP
asymmetry in $B_d^0 \to \phi K_s$ which originates from the
$\tilde{b}_R-\tilde{s}_R$ mixing should be generically suppressed.
This argument leaves $\tilde{b}_L-\tilde{s}_L$ mixing as a possible
source of large CP asymmetry in $b \to s$ transition.  Thus it is
important to investigate to what extent large
$\tilde{b}_L-\tilde{s}_L$ mixing is allowed.

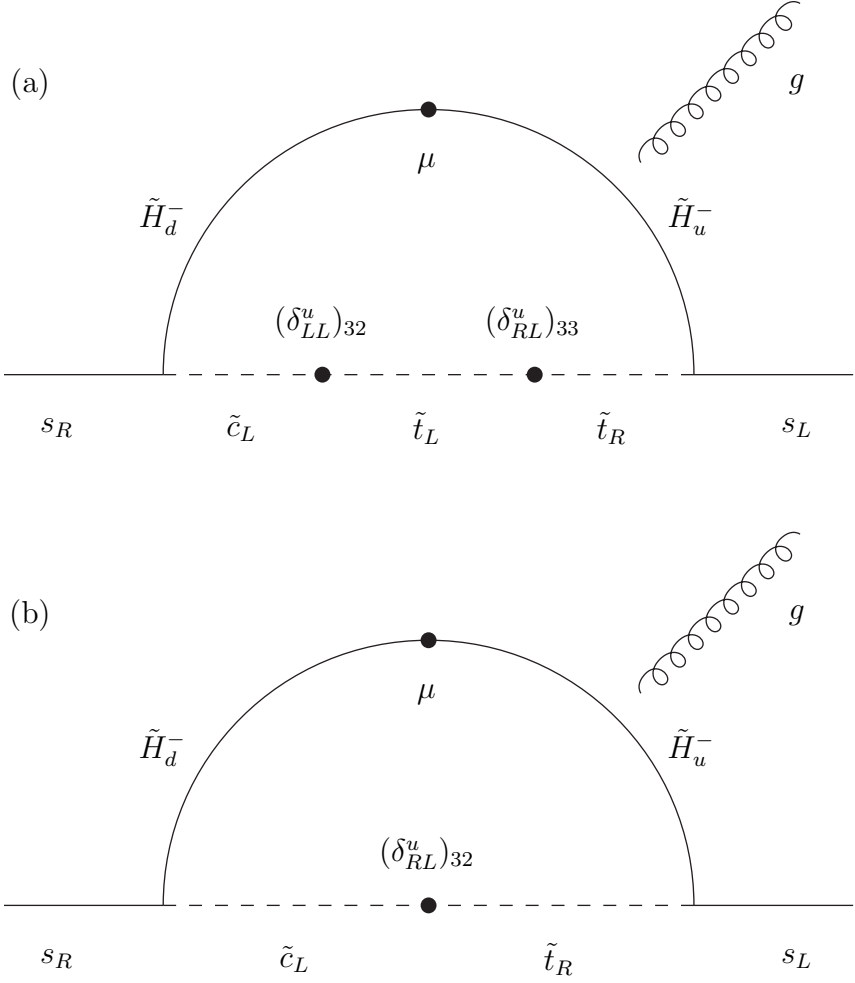
\begin{figure}[h]
  \begin{center}
    \begin{picture}(400,200)(0,0) 
      \Text(50,150)[]{(a)}
      \Line(40,40)(100,40) \DashLine(100,40)(300,40){5} \Line(300,40)(360,40)
      \Text(60,20)[]{$s_R$}
      \Text(130,20)[]{$\tilde{c}_L$}
      \Text(200,20)[]{$\tilde{t}_L$}
      \Text(270,20)[]{$\tilde{t}_R$}
      \Text(340,20)[]{$s_L$}
      \Vertex(160,40){3} \Text(160,60)[]{$(\delta_{LL}^u)_{32}$}
      \Vertex(240,40){3} \Text(240,60)[]{$(\delta_{RL}^u)_{33}$}
      \Vertex(200,140){3} \Text(200,120)[]{$\mu$}
      \CArc(200,40)(100,0,180)
      \Text(100,100)[]{$\tilde{H}_d^-$} \Text(300,100)[]{$\tilde{H}_u^-$}
      \Gluon(280,120)(340,180){4}{8} \Text(340,150)[]{$g$}
    \end{picture}
    \begin{picture}(400,200)(0,0) 
      \Text(50,150)[]{(b)}
      \Line(40,40)(100,40) \DashLine(100,40)(300,40){5} \Line(300,40)(360,40)
      \Text(60,20)[]{$s_R$}
      \Text(150,20)[]{$\tilde{c}_L$} \Text(250,20)[]{$\tilde{t}_R$}
      \Text(340,20)[]{$s_L$}
      \Vertex(200,40){3} \Text(200,60)[]{$(\delta_{RL}^u)_{32}$}
      \Vertex(200,140){3} \Text(200,120)[]{$\mu$}
      \CArc(200,40)(100,0,180)
      \Text(100,100)[]{$\tilde{H}_d^-$} \Text(300,100)[]{$\tilde{H}_u^-$}
      \Gluon(280,120)(340,180){4}{8} \Text(340,150)[]{$g$}
    \end{picture}
    \caption{The dominant Feynman diagrams which contribute to the CEDM 
      of the strange quark via chargino exchange.}
    \label{fig:diagram}
  \end{center}
\end{figure}

\begin{figure}[h]
  \begin{center}
    \includegraphics[scale=0.8]{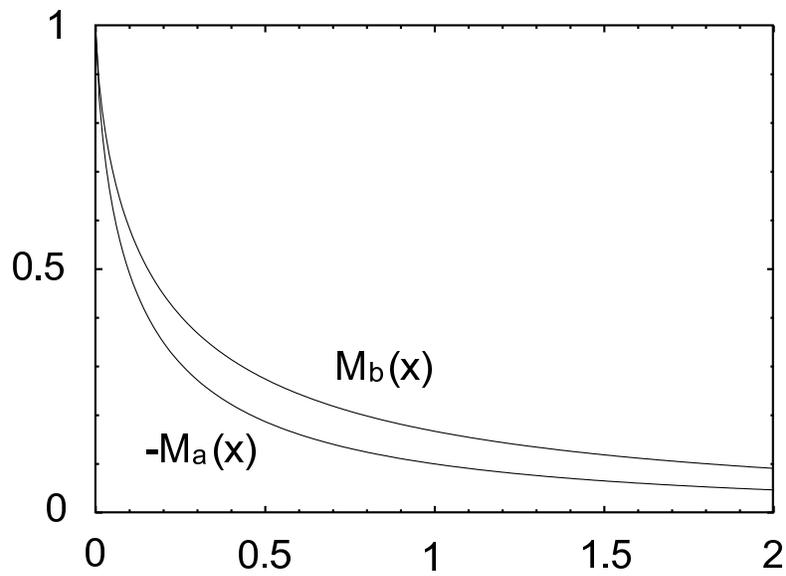}
  \end{center}
  \caption{The dependence of the loop functions $M_a$ and $M_b$ on 
    $x = |\mu|^2/m_{\tilde{q}}^2$. 
    Notice that $-M_a$ is plotted instead of $M_a$.}
  \label{fig:Mab}
\end{figure}

We now point out that the LL mixing is
also constrained by the $^{199}$Hg EDM bound.
We first emphasize that chargino-mediated processes can also 
provide large contributions to the CEDM of the strange quark $d_s^C$
because there are diagrams which are enhanced by the top Yukawa coupling 
constant (See Fig. \ref{fig:diagram}).
In the presence of $2-3$ mixing in the LL sector,
$d^C_s$ is induced by the double mass insertion diagram in 
Fig. \ref{fig:diagram}(a) and is evaluated as
\footnote{It was shown that the QCD running effect 
in the CEDM operator Eq.(\ref{eq:CEDM_operator})
from the weak scale to the hadronic scale is small \cite{Falk:1999tm}.
In fact, the enhancement factor is calculated as $c \sim 0.9$, 
and thus we neglect it in our analysis.}
\begin{eqnarray}
  d_s^C &=& \frac{1}{32 \pi^2} y_t y_s V_{ts} 
  \frac{1}{m^2_{\tilde{q}}} 
  M_a(x) \mbox{Im}  [\mu (\delta_{RL}^u)_{33} (\delta_{LL}^u)_{32}]
  \nonumber \\
  &\simeq& \frac{1}{8 \pi^2} \frac{G_F}{\sqrt{2}}  
  m_t^2 m_s V_{ts}   \frac{|A_t \mu| \tan \beta}{m_{\tilde{q}}^4} 
  M_a(x) |(\delta_{LL}^u)_{32}| \sin \theta_a,
  \label{eq:cedm}
\end{eqnarray}
by using the so-called mass insertion method.
Here the loop function $M_a(x)$ is defined as
\begin{eqnarray}
  M_a(x) & \equiv & 
  \frac{1 + 9x - 9x^2 - x^3 + 6x(1+x) \log x}{(x-1)^5},
\end{eqnarray}
where $x \equiv {|\mu|^2}/{m_{\tilde{q}}^2}$.
One observes that $M_a(1) = -0.1$, $M_a(0.25) \simeq -0.31$ and $M_a(0) = -1$.
The behavior of the function $M_a(x)$ is depicted in Fig. \ref{fig:Mab}.
Here the mass insertion parameter $(\delta_{RL}^u)_{33}$ is given as
\begin{eqnarray}
  (\delta_{RL}^u)_{33} = 
  - \frac{m_t (A_t + \mu^* \cot \beta)}{m_{\tilde{q}}^2}
\end{eqnarray}
and the term proportional to $\cot \beta$ can be neglected.
$\theta_a$ parameterizes the CP violating phase as
$\theta_a = \arg [A_t \mu (\delta_{LL}^u)_{32}]$.
Thus $d_s^C$ from Fig. \ref{fig:diagram}(a) is estimated as
\begin{eqnarray}
  e|d_s^C| \simeq && 2.8 \times 10^{-24}\ e\ \mbox{cm} 
  \times |(\delta_{LL}^u)_{32} \sin \theta_a |
  \nonumber \\
  && \times 
  \left( \frac{\tan \beta}{20} \right)
  \left( \frac{|\mu|}{250~\mbox{GeV}} \right)
  \left( \frac{|A_t|}{500~\mbox{GeV}} \right)
  \left( \frac{m_{\tilde{q}}}{500~\mbox{GeV}} \right)^{-4}
  \left( \frac{M_a(x)}{-0.31} \right).
\end{eqnarray}
Compared with the experimental upper limit Eq. (\ref{eq:bound-on-cedm-of-s}), 
we obtain
\begin{eqnarray}
  |(\delta_{LL}^u)_{32} \sin \theta_a| < 0.20
  \label{eq:constraint_on_LL}
\end{eqnarray}
for $\tan\beta = 20$, $|\mu| = 250\ {\rm GeV}$ and
$|A_t| = m_{\tilde{q}} = 500\ {\rm GeV}$.
Therefore we conclude that the $\tilde{t}_L-\tilde{c}_L$ mixing angle or the 
CP violating phase must be suppressed at the level of $O(0.1)$ when 
$\tan \beta$ is large, in the light of the result of the  $^{199}$Hg EDM 
experiment.

It is important to note  
that $\tilde{t}_R-\tilde{c}_L$ mixing is induced 
by a combination of $\tilde{t}_R-\tilde{t}_L$ and $\tilde{t}_L-\tilde{c}_L$
mixing through the double mass insertion diagram:
\begin{eqnarray}
  (\delta_{RL}^u)_{32}^{\rm ind} =
  - \frac{m_t (A_t + \mu^* \cot \beta)}{m_{\tilde{q}}^2} \times 
  (\delta_{LL}^u)_{32}.
\end{eqnarray}
Based on this observation,
it is manifest that $(\delta_{RL}^u)_{32}$ is also constrained considering
the diagram in Fig. \ref{fig:diagram}(b).
$d_s^C$ from this contribution is given by
\begin{eqnarray}
  d_s^C &\simeq& \frac{1}{8 \pi^2} \frac{G_F}{\sqrt{2}}  
  m_t m_s V_{ts} \frac{|\mu| \tan \beta}{m_{\tilde{q}}^2} 
  M_b(x) |(\delta_{RL}^u)_{32}| \sin \theta_b
\end{eqnarray}
where
\begin{eqnarray}
  M_b(x) \equiv  \frac{1 + 4x - 5x^2 + 2x(2+x) \log x}{(x-1)^4}
\end{eqnarray}
(See Fig. \ref{fig:Mab}) and $ \theta_b = \arg [\mu (\delta_{RL}^u)_{32}]$.
Following the same procedures, 
the $\tilde{t}_R-\tilde{c}_L$ mixing parameter is found to be 
constrained as 
\begin{eqnarray}
  |(\delta_{LR}^u)_{23} \sin \theta_b| < 5.5 \times 10^{-2}
  \label{eq:constraint_on_LR}
\end{eqnarray}
for the same SUSY parameters as in the $\tilde{t}_L-\tilde{c}_L$ case.

We stress that 
the constraint on the mixing parameter $(\delta_{LL}^u)_{32}$ given 
by Eq. (\ref{eq:constraint_on_LL}) is 
severer than that obtained from 
the branching fraction of the process $b \to s \gamma$.
So far the limit on $b \to s \gamma$ was considered to give the 
strongest bound on $(\delta_{LL}^u)_{32}$ taking chargino-loop diagrams
into consideration \cite{Causse:2000xv}.
However we find that the $^{199}$Hg EDM generically provides severer
constraint on $(\delta_{LL}^u)_{32}$ by about one or two orders of magnitude
depending on mass spectra of superparticles, provided that the CP phase is 
of order unity.  The same argument is also applied to $(\delta_{LR}^u)_{23}$.

We have dealt with the diagrams shown in Fig. \ref{fig:diagram} exclusively.
Let us consider other possible diagrams 
producing the CEDM of the strange quark.
For example, the constraint on the mixing 
$(\delta^u_{LL})_{31}$ is turned out to be weakened 
by the Cabibbo angle and that on 
the RR mixing $(\delta^u_{RR})_{32}$ by about $m_c/m_t$ compared with that on
$(\delta^u_{LL})_{32}$.
Thus we find that the LL mixing $(\delta^u_{LL})_{32}$ most 
considerably enhances 
the chargino-origin dipole moment of the strange quark 
among off-diagonal elements of the scaler quarks and thus
should be bounded most stringently.

Notice that 
$(\delta^u_{LL})_{ij}$ and $(\delta^d_{LL})_{ij}$ are related to each other
because of the $SU(2)_L$ symmetry.
In the superCKM basis, one finds that
\begin{equation}
    (\delta^d_{LL})_{ij}=\sum_{k,l}(V^*_{\rm{CKM}})_{ki}
              (\delta^u_{LL})_{kl} (V_{\rm{CKM}})_{lj},
\end{equation} 
where $V_{\rm{CKM}}$ is the Cabibbo-Kobayashi-Maskawa (CKM) matrix. 
The $(3,2)$ component can be expanded as
\begin{equation}
(\delta^d_{LL})_{32} \sim (\delta^u_{LL})_{32} 
+ \lambda (\delta^u_{LL})_{31} + O(\lambda^2)
\end{equation}
in terms of the Wolfenstein parameter $\lambda \sim 0.2$.
Thus in the absence of accidental cancellation or hierarchy
among the parameters
the bound on $(\delta^d_{LL})_{32}$ is translated into that on 
$(\delta^u_{LL})_{32}$, and vice versa.
The severest bound on $(\delta^d_{LL})_{32}$ comes from
gluino-mediated $b \to s \gamma$ via a double mass insertion 
\cite{Causse:2000xv}.
We stress that the constraint on the $2-3$ mixing in the LL sector
from the $^{199}$Hg EDM is
generically as strict as that in this $b \to s \gamma$ process,
though detailed comparison is sensitive to the SUSY mass spectra.\footnote{
This argument is applied to the generic case where there is no accidental 
cancellation or no hierarchy among the squark flavor mixing parameters so that $(\delta^d_{LL})_{32} \sim (\delta^u_{LL})_{32}$.  
On the other hand, the constraint on the mixing of the up-type squarks 
Eq.(\ref{eq:constraint_on_LL}) obtained in a previous paragraph is very
general.}

Let us now discuss the $B_d^0 \to \phi K_s$ process, which is known to
be one of the most interesting modes. The Belle collaboration \cite{Belle} 
has reported 
the deviation of the CP asymmetry in $B_d^0 \to \phi K_s$ from the SM 
prediction, while the latest result from the BaBar collaboration \cite{BaBar} 
is 
consistent with the SM prediction. From the theoretical side, CP violation 
additional to the SM is expected to manifest itself in this process because
the tree-level SM contribution is absent.
Although it is shown that both $\tilde{b}_L-\tilde{s}_L$ and
$\tilde{b}_R-\tilde{s}_R$ mixing should be suppressed because of the
$b \to s \gamma$ and the $^{199}$Hg EDM constraints, there still
remains some room to cause a sizable effect on the indirect CP-violation 
parameter $S_{\phi K_S}$.  Notice
that the bound on the $\tilde{b}_L-\tilde{s}_L$ mixing parameter is
not as tight as that on $\tilde{b}_R-\tilde{s}_R$.  We emphasize that
the prediction of $S_{\phi K_S}$ can be considerably affected since
$\tilde{b}_L-\tilde{s}_L$ mixing combines with
$\tilde{b}_R-\tilde{b}_L$ mixing to induce significant
$\tilde{b}_R-\tilde{s}_L$ mixing for large $\tan \beta$
\cite{Khalil:2002fm,Kane:2002sp,Harnik:2002vs}.  On the contrary, the mass
difference between $B_s$ and $\bar{B}_s$ would not deviate
considerably from its SM prediction under the circumstance and is
expected to be measured at Tevatron Run II
\cite{Kane:2002sp,Harnik:2002vs}.  While the contribution from the induced
$\tilde{b}_R-\tilde{s}_L$ mixing is enhanced in the dipole operators
and thus considerably affects $b \to s \gamma$ and $B_d^0 \to \phi
K_s$ processes, box diagrams which involve the induced
$\tilde{b}_R-\tilde{s}_L$ mixing are sub-dominant in $B_s-\bar{B}_s$
mixing.

Here we will discuss loopholes that escape the new bound given above, namely possibilities of cancellation or suppression 
of the contributions to the $^{199}$Hg EDM:
(i) As masses of superparticles increase, 
the contribution to the dipole moment becomes suppressed;
(ii) There might be the case where
various contributions to the $^{199}$Hg EDM cancel out each other.  
The gluino-loop contribution may cancel
out the chargino contribution discussed above.
Also, though we have considered the contribution 
only from the strange quark, 
it may be possible that the contributions 
from the up and down quarks compensate it and that
the resulting mercury EDM becomes negligible;
(iii) The phase of $\mu A_t$ may align with that of 
$(\delta_{LL}^u)_{32}$ by some mechanism.  
In fact, the imaginary part of $\mu A_t$
is less constrained by EDM experiments;
(iv) There remains theoretical uncertainty in evaluating the EDM of $^{199}$Hg
atom, which relies on calculations on nuclear and hadron dynamics.

To conclude, we would like to emphasize the importance 
of constraining the CP-violating $2-3$ mixing in the LL sector.
Given the minimal supergravity boundary condition, 
$(\delta^d_{LL})_{32}$ emerges from renormalization group running
and becomes proportional to $V_{ts}$, and thus its 
imaginary part is naturally suppressed
since there arises no new CP-violating phase except for the CKM matrix.
However in the case where the solution of the SUSY flavor problem is 
attributed to some alignment mechanism based on flavor symmetries
generation mixing is expected to be accompanied by 
large CP-violating phases independent of the CKM matrix.
Therefore our new constraint on CP-violating mixing may offer a crucial hint
in building models of flavor, which will be discussed elsewhere \cite{Endo-Kakizaki-Yamaguchi}.

\section*{Acknowledgment}
This work was supported in part by the Grants-in-aid from the Ministry
of Education, Culture, Sports, Science and Technology, Japan, No.12047201 and
No.14046201.
ME and MK thank the Japan Society for the Promotion of Science for financial 
support.


\end{document}